\documentclass[sigconf]{acmart}
\AtBeginDocument{%
  }

\copyrightyear{2025}
\acmYear{2025}
\setcopyright{cc}
\setcctype{by}
\acmConference[C\&C '25]{Creativity and Cognition}{June 23--25, 2025}{Virtual, United Kingdom}
\acmBooktitle{Creativity and Cognition (C\&C '25), June 23--25, 2025, Virtual, United Kingdom}\acmDOI{10.1145/3698061.3726910}
\acmISBN{979-8-4007-1289-0/2025/06}

\usepackage{multirow} 
\usepackage{xcolor}

\newif\ifsubmission
\submissiontrue

\ifsubmission
  \newcommand{\added}[1]{#1}
  \newcommand{\deleted}[1]{} 
\else
  \newcommand{\added}[1]{\textcolor{blue}{#1}} 
  \newcommand{\deleted}[1]{\textcolor{red}{#1}} 
\fi

\begin{document}

\title[From Pen to Prompt: How Creative Writers Integrate AI into their Writing Practice]{From Pen to Prompt: How Creative Writers Integrate AI into their Writing Practice}

\author{Alicia Guo}
\email{axguo@cs.washington.edu}
\affiliation{%
  \institution{University of Washington}
  \city{Seattle}
  \country{United States}
}

\author{Shreya Sathyanarayanan}
\email{sathshr@cs.washington.edu}
\affiliation{%
  \institution{University of Washington}
  \city{Seattle}
  \country{United States}
}

\author{Leijie Wang}
\email{leijiew@cs.washington.edu}
\affiliation{%
  \institution{University of Washington}
  \city{Seattle}
  \country{United States}
}

\author{Jeffrey Heer}
\email{jheer@uw.edu}
\affiliation{%
  \institution{University of Washington}
  \city{Seattle}
  \country{United States}
}

\author{Amy Zhang}
\email{axz@cs.uw.edu}
\affiliation{%
  \institution{University of Washington}
  \city{Seattle}
  \country{United States}
}

\renewcommand{\shortauthors}{Guo et al.}



\begin{abstract}
Creative writing is a deeply human craft, yet AI systems using large language models (LLMs) offer the automation of significant parts of the writing process. So why do some creative writers choose to use AI? Through interviews and observed writing sessions with 18 creative writers who already use AI regularly in their writing practice, we find that creative writers are intentional about how they incorporate AI, making many deliberate decisions about when and how to engage AI based on their core values, such as authenticity and craftsmanship. We characterize the interplay between writers' values, their fluid relationships with AI, and specific integration strategies---ultimately enabling writers to create new AI workflows without compromising their creative values. We provide insight for writing communities, AI developers and future researchers on the importance of supporting transparency of these emerging writing processes and rethinking what AI features can best serve writers.
\end{abstract}



\maketitle







\section{Introduction}

\begin{quote}
``\textit{...[A]rt is something that results from making a lot of choices... When you are writing fiction, you are---consciously or unconsciously---making a choice about almost every word you type; to oversimplify, we can imagine that a ten-thousand-word short story requires something on the order of ten thousand choices. When you give a generative-A.I. program a prompt, you are making very few choices; if you supply a hundred-word prompt, you have made on the order of a hundred choices}.'' \\
\null\hfill  --- Ted Chiang, ``Why A.I. Isn't Going to Make Art'' \cite{Chiang2024WhyAI}
\end{quote}

Creative writing, whether in fiction, poetry, or experimental forms, is regarded as a deeply personal and human craft grounded in artistic expression. Can it be compatible with new large language model (LLM)-based technologies that offer to automate significant portions of the writing process? While computational support for creative writing has existed for decades, rapid improvements in text generation brought by LLMs raise new concerns about the impact of these tools on the human elements of creative work and writers' labor. Some concerns are broadly shared, such as societal harm from AI due to lack of attribution \cite{huang2024AuthorshipAttribution}, hallucinations \cite{Wei2024DetectingHallucinations}, ethical considerations \cite{McKee2020EthicsAIWriting}, and job displacement \cite{wang2023LLMParadox}.

But the unique properties of \textit{creative} writing---as opposed to academic, business, journalistic, or marketing writing, where LLMs have proven to help with productivity \cite{Jin2024ACS, Zhang2021AutomaticPC, jovic2024AI-GeneratedEmails, zhou2024correcting}---raise additional concerns about how AI usage can impact aspects of one's writing \textit{craft}, like authenticity, ownership, and the creative process \cite{hwang2024it80me20, gero2023social}. For some, AI feels fundamentally incompatible with the creative act, reducing a multitude of creative choices to a few prompts \cite{Chiang2024WhyAI}. Others view AI as potentially another tool in a writer's creative arsenal, akin to digital word processors and editing software \cite{biermann2022fromtool}. This tension is evident as literary magazines and creative writing platforms have struggled to decide on policies for AI-assisted work amidst strong disagreements in the creative writing community \cite{NaNoWriMo2024}. 

Previous work has explored writers' hypothetical responses to AI and studied the use of specific tools through lab experiments \cite{novelists, Chakrabarty2024Creativity, elephant, ippolito2022creative}. \added{Many of these studies primarily focus on isolated writing tasks or specific AI tools in controlled settings, rather than observing writers' complete practices across multiple platforms. Additionally, prior studies often involve participants who are new to AI writing tools, capturing initial reactions rather than evolved practices. This prior research establishes important foundations regarding writers' initial concerns and challenges with AI tools}---however, few studies have examined the long-term qualitative experience of creative writers who have \textit{actively chosen} to integrate AI into their workflows and its effects on both their craft and career. \added{This experience becomes especially important as research has shown that AI tools become more useful over time as users familiarize themselves \cite{long}}. These writers suggest that there may be ways to preserve essential human elements of creative writing while embracing new technological possibilities. 

To understand how creative writers view and work with AI over the long term, we conducted a study using interviews and contextual inquiry with 18 creative writers who have integrated language-based generative AI (for months or years) into their writing process to consider the questions:

\begin{itemize}
    \item RQ1: What drives creative writers to incorporate AI? When and why do they choose to use AI?
    \item RQ2: What strategies and workflows emerge as writers develop personal practices for writing with AI?
\end{itemize}

Our participants represent a distinct group: creative writers who have chosen to use AI despite potential concerns. In this moment where the role of AI in creative work is still evolving, these writers are uniquely positioned to reflect on their experiences before and after writing with AI. Having used AI long enough to develop their own practices, they shed light on sustainable approaches to AI integration in creative practice. 
Through interviews and direct observation of their writing processes, we examine their strategies used to navigate tensions around AI and the emergent practices that \added{only arise over continued use.} 

We find that creative writers actively negotiate between AI's benefits and their deeply held writerly values, setting boundaries around AI use to preserve love of the craft. Rather than viewing AI simply as tool or collaborator, writers develop fluid relationships with AI that shift with their goals, influencing and reinforcing AI use. These relationships are supported by sophisticated workflows involving many iterations and decision points, maintaining creative control through careful decision-making at every stage. We characterize this dynamic interplay of values, relationships, and strategies to reveal how different approaches to AI emerge across individuals and genres. Our findings provide insight for writing communities, tool designers, and future research. As the landscape evolves, understanding how writers maintain their craft and values while embracing technological support will be crucial for developing AI and practices that support instead of undermine human creativity.


\section{Related Work}


To understand why and how creative writers are integrating AI into their practice, we look to current issues with AI in creative writing communities, along with previous work in computational support for creative writing and studies of  writer-AI interactions.

\subsection{Effects of AI on creative writing communities}



Writers are concerned about the artistic integrity of AI use in writing, where critics see it as ``pushing a button'' to generate books \cite{Chiang2024WhyAI}, thereby devaluing human labor and creativity. Studies also show people are generally biased against AI-assisted works \cite{horton_jr_bias_2023, hong}. A 2022 controversy over an AI-generated artwork winning a major prize raised concerns about the authenticity and effort behind AI-driven creations \cite{Roose2022artprize}, echoing broader challenges that AI presents to traditional concepts of authorship, ownership, and the creative process \cite{hertzmann2018computerscreateart}. For many creatives, AI raises concerns about public distrust \cite{kawakami}, economic loss  \cite{jiang}, and labor replacement \cite{revolution, Marr2024AIChangeJobs}---issues that contributed to the 2023 Writers Guild of America strike where screenwriters fought to limit the use of AI in their work. 

While AI benefits individual writers by helping them be more productive and creative \cite{doshi, Noy2023ExperimentalEO}, studies also suggest AI can lead to creative homogenization \cite{homogenization, li2024, doshi}. Self-publishing platforms have seen a surge in AI-assisted work \cite{ScifiNPR, sato}, simultaneously raising concerns about market saturation and spam. Meanwhile, tools like ``Have I Been Trained'' \cite{HaveIBeenTrained} showing how writers' works are being trained on without consent have fueled demands for better frameworks around data rights, compensation, and legal clarity around copyright. Creative institutions are still adapting---some magazines temporarily halted submissions after being inundated with AI generated content largely from outside the writing community \cite{ScifiNPR}, and many venues have introduced explicit policies banning or requiring disclosure of AI use. The decision of NaNoWriMo (a popular month-long novel writing challenge) to allow AI use faced intense backlash, highlighting divisions within creative writing communities \cite{NaNoWriMo2024}.

\subsection{Computational support for creative writing}


Computational tools have supported creative writing for decades, from rule-based systems \cite{Howe2009RiTaCS, Compton2015TraceryAA} to chatbots. Over time, computational support evolved to address almost every stage of the writing process from ideation \cite{shaer2024AI-Augmented,Han2018Combinator, secondmind}  to rewriting and feedback \cite{chakrabarty, zhou2024Aillude} across creative genres ranging from storytelling \cite{kreminski} to poetry \cite{gero2019Metaphoria} to theater scripts \cite{screenplay}. Modern AI writing systems offer integration at multiple stages of the writing process. For example, commercial AI platforms made for creative writers, such as Sudowrite \cite{Sudowrite} and Novelcrafter \cite{novelcrafter}  generate inline text continuations that maintain stylistic and narrative consistency alongside more focused features such as structured story outlining and worldbuilding assistance. Chat-based platforms like ChatGPT \cite{chatgpt} let writers explicitly prompt for creative writing support. Other tools provide support for focused tasks, such as generating customized thesauruses \cite{thesaurus}, suggesting connections for new metaphors \cite{gero2019Metaphoria},  and building characters through chat \cite{charactermeet} or simulation \cite{kreminski}.

\subsection{Understanding writer interactions with AI}


A growing body of work has characterized writer perceptions and usage of AI, with writers finding AI helpful in the planning, translating, and reviewing stages of the writing process \cite{novelists, Chakrabarty2024Creativity, elephant, personas}; however, much of this work uses controlled lab studies with custom AI interfaces. Differences between writers emerged over longer settings---when asked to use a custom AI interface over an 8-week period, researchers observed some choosing to incorporate AI text directly while others chose to use AI primarily for ideation and rewriting \cite{ippolito2022creative}.
Studies also observed differences between types of writers. Novices perceive greater benefits from using AI for ideation while experienced writers struggle with aligning AI-generated content to their vision \cite{buncho, ippolito2022creative}. Hobbyists emphasize personal expression and ownership, while professionals prioritize productivity \cite{biermann2022fromtool}. 
Indeed, in broader creative domains, researchers have seen a shift in creative professionals' work towards high-level orchestration much like that of a project managers' workflow \cite{evolving}.

Relatively less research has examined interaction paradigms between writers and AI. Work characterizing writing support tools found that they can be categorized anywhere from cognitive aids to collaborators \cite{gero2019Metaphoria, Compton2015CasualC, catherine2024GhostWriter, Yang2022AIAA, redhead}. 
To understand the interplay between writer and AI, researchers have built visualizations and analyses of log data capturing writer interactions with AI from crowdsourced participants \cite{lee2022CoAuthor, coauthorviz}.
  As creative writers interact with AI tools, common challenges include preserving narrative coherence \cite{ippolito2022creative, buncho}, maintaining stylistic voice \cite{GhostwriterEffect, ippolito2022creative}, diminishing senses of ownership and satisfaction \cite{hwang2024it80me20, whatuse}, and worries about plagiarism \cite{intelligentsupport}. 
Moreover, depending on its use, AI assistance does not necessarily improve text quality \cite{intelligentsupport, slogan}. 
Finally, social dynamics impact when writers might prefer AI support to a human peer, with factors including writers' desires for support, perceptions of human or AI support, and values about writing \cite{gero2023social}. 

This work builds upon these foundations by examining writers who have \textit{chosen} to incorporate AI into their creative practice and have had months to years to develop their workflows. Previous approaches often fragment the creative process by focusing on specific stages or tools, with participants who are either new to AI for creative writing or the tools they are asked to use. Research has shown that AI tools become more useful over time \added{as users learn and familiarize themselves with the system, move past the novelty effect, and begin customizing and appropriating the technology} \cite{long, novelty}, highlighting the importance of sustained engagement with tools rather than exploratory use. Rather than studying specific tools or writing tasks in isolation, we observe writers' complete workflows \textit{across} multiple commercial AI writing platforms from different genres. By combining interviews with direct observation of writers' practices, we capture the concrete decisions and \added{emergent patterns that only develop} \textit{in practice} as writers establish sustainable approaches to working with AI \added{past the familiarization phase.}

\section{Methods}

Over the course of November 2023 to July 2024, the first author conducted 90-minute semi-structured interviews virtually with the writers, spending \added{around} 30 minutes of the interview time directly observing their writing process.

\subsection{Recruitment and Participants}

We focused on writers in fiction, creative non-fiction, and poetry, while allowing participants to self-identify as creative writers without strict requirements for publications or formal education. We recruited writers from AI-focused online forums and communities, university writing and MFA email lists, and used snowball sampling to expand our participant pool. Participants completed a screening survey covering their writing and AI experience, and we selected writers who demonstrated experience with AI writing tools, either through sustained use (at least a month) or completed works.

The final participants included 18 creative writers in different genres, with fiction most represented (Table \ref{tab:ai_writing_participants}). 
Writers' experience with AI span months to years, and they use a variety of AI tools from chatbots to more integrated inline-generation platforms with many focused features. Writers' reported use of AI is detailed in Appendix \ref{sec:AIusesurvey}.
Nine had no formal writing education, 3 had masters/grad degrees in writing, and 6 took writing classes in college or online. Twelve participants resided in the USA, 5 in Europe, 1 in Canada, and 1 undisclosed. Ten participants identified as female, 6  as male, 1 as non-binary, and 1 undisclosed.

\begin{table*}[ht!]
\caption{Summary of participant demographics, their writing background, and experience writing with AI.}
\small
\centering
\begin{tabular}{|p{1.3em}|p{8em}|p{13em}|p{3.9em}|p{3.7em}|p{14em}|p{2.5em}|} \hline
\textbf{ID}& \textbf{Published} & \textbf{Genres with AI} & \textbf{Years writing {\scriptsize (in general)}} & \textbf{Years writing {\scriptsize (with AI)}} & \textbf{AI used} & \textbf{Age} \\
\hline
P1& Yes & Fiction (Novels) & \mbox{>~10~years} & 5 months & ChatGPT, Claude, Sudowrite & 65+ \\ \hline 
P2& Self Published & Fiction (Novels) & 9 years & 2 years & ChatGPT, Claude, Novelcrafter & 46-55 \\ \hline 
P3& Yes & Fiction (Short Stories) & 4 years & 2 years & ChatGPT, Claude & 18-25 \\ \hline 
P4& Personal Blog & Personal Essays & 3 years & <~1 year & ChatGPT & 18-25 \\ \hline 
P5& Shared Online & Fiction (Short Stories, Fanfiction), Poetry & 4-5 years & \mbox{\textasciitilde{} 1 year} & \scriptsize ChatGPT, Claude, code-davinci-002 and davinci-002 base models through Obsidian plugin "Loomsidian" & 18-25 \\ \hline 
P6& N/A & Fiction (Short Stories) & \mbox{>~10~years} & 6 months & NovelAI, Claude, ChatGPT  & 26-35 \\ \hline 
P7& Self Published & Fiction (Novels, Short Stories) & \mbox{\textasciitilde{}10 years} & 1.5 years & ChatGPT, NovelAI & 26-35 \\ \hline 
P8& Personal Blog & Personal Essays & n/a & \mbox{\textasciitilde{}1 year} & ChatGPT, Claude & n/a \\ \hline 
P9& N/A & Poetry, Short Stories & 4 years & 2.5 years & ChatGPT, Sudowrite, Marlowe & 18-25 \\ \hline 
P10& Self Published & Fiction (Novels, Novellas) & \mbox{>~10~years}  & 2 years & ChatGPT, NovelCrafter, Perplexity & 65+ \\ \hline 
P11& Yes & Fiction (Experimental, Short Stories) & \mbox{>~10~years} & 2.5 years & Write With LAIKA, Pi, Midjourney, Runway & 36-45 \\ \hline 
P12& Self Published & Fiction  & \mbox{>~10~years} & 2 years & ChatGPT, Claude, NovelCrafter & 26-35 \\ \hline 
P13& Personal Blog & Poetry, Lyrics, Personal Essays & \mbox{>~10~years} & 1.5 years & ChatGPT & 26-35 \\ \hline 
P14& Self Published  & Fiction (Novels, Novellas, Short Stories) & \mbox{>~10~years} & 3 years & ChatGPT, Claude, Rexy & 36-45 \\ \hline 
P15& Literary Magazines & Poetry & \mbox{>~5 years} & 1.5 years & ChatGPT & 26-35 \\ \hline 
P16& Literary Magazines & Poetry & 8 years & \mbox{\textasciitilde{}2 years} & ChatGPT, Copilot, Bard & 26-35 \\ \hline 
P17& Personal Blog & Lyrics & 3-4 years & 1.5 years & ChatGPT & 18-25 \\ \hline 
P18& Literary Magazines & Poetry, Prose & \mbox{>~10~years} & \mbox{\textasciitilde{}1 year} & ChatGPT, Claude, Bard, Gemini & 46-55 \\ \hline
\end{tabular}

\label{tab:ai_writing_participants}
\end{table*}

\subsection{Interview Procedure}
The 90 minute semi-structured interviews were conducted over Zoom. Participants were asked about their writing backgrounds and experiences to provide context for their creative processes. Questions then explored how they engage with AI tools in their writing, focusing on their motivations, workflows, and decision-making. Follow-up questions aimed to understand broader themes such as the role of AI in creativity, ethical considerations, and perceptions of originality and authorship. 

Interviews included an \textit{observed live-writing session} lasting approximately 30 minutes \added{(range: 29-45 minutes, mean: 34.5 minutes),} where writers shared their screen and worked on a current writing project \added{(novel chapters, poems, essays)} using their preferred AI platforms (see Appendix \ref{sec:writingsession} for details on participants' writing projects and setups). \added{Participants were instructed to write naturally as they would in their normal practice, without any constraints on their process or AI usage and encouraged to think aloud about their decision-making to the degree they felt comfortable. The researcher asked contextual questions about specific actions such as ``\textit{what made you accept this suggestion?}'' but otherwise minimized intervention to preserve the authenticity of their process.} Writing while being observed can be a potentially uncomfortable position. To minimize observer effects, participants were given alternatives to live observation: 1) prerecording their writing process before the interview, or 2) keeping a diary of their writing process before the interview. One participant \added{chose option 1 and} prerecorded themselves writing, then participated in the interview where the writer asked questions about their recording in place of the observation session. The final interview guideline is detailed in Appendix \ref{sec:interview_questions}. This study was reviewed by our institution's IRB and deemed exempt.


\subsection{Data Analysis}

We analyzed interview and writing session data following grounded theory methods \cite{MullerKogan2012}. After data collection, \added{the interview recordings were automatically transcribed and aligned with the videos, then manually corrected for errors.} Initial codes were developed by the first and third author, then updated with the second author as new interviews were conducted. We additionally coded non-verbal interactions (e.g., prompting method)  from the screen recordings that transcripts couldn't capture. The authors discussed emerging themes after every few interviews and updated the interview guideline accordingly. The data was open-coded by the authors and clustered into themes, with the first author leading the process. 



\section{Findings}

We identified three key elements that shape how writers engage with AI: their personal \textit{values} about writing, their perceived \textit{relationships} with AI, and their concrete AI \textit{integration strategies}. Rather than operate independently, these elements form a dynamic system where each element influences and reinforces the others. For example, writers who strongly value ownership will often relate to AI as an assistant or tool, in order to maintain explicit control. These relationships in turn shape what integration strategies writers develop: writers viewing AI as an assistant typically develop more bounded, task-specific workflows compared to those who see AI as a collaborator. \deleted{However, influence also flows in the other direction, where writers' experiences with using AI can shift how they view their relationship over time.}

Through these findings, we characterize the interactions of values, relationships, and integration strategies to describe the factors that shape how writers approach AI, providing a foundation for writers' decisions. The themes that follow demonstrate \textit{when} and \textit{why} writers choose to use AI, to then better understand \textit{how} writers are integrating and making sense of AI in their workflows. 

\subsection{When and Why Do Creative Writers Use (and Not Use) AI?}


We found that values play a major role in when and why writers choose to use AI, setting usage boundaries to preserve what matters most in their creative practice. To ground our findings, we begin by briefly summarizing the core \textit{writers’ values} that emerged:

\begin{itemize}
  \item \textbf{Authenticity}: maintaining their unique voice and expression
  \item\textbf{Ownership}: feeling that the text is theirs and having control over the creation process
  \item\textbf{Creativity}: expressing themselves in distinctive ways
  \item\textbf{Craftsmanship}: intentionality, effort, and quality
\end{itemize}

A fuller description of these values and example quotes can be found in Appendix \ref{sec:writers_values}. While prior work has explored taxonomies of writers' values and writer challenges that overlap with our findings \cite{ippolito2022creative, gero2023social, biermann2022fromtool}, the goal of our inquiry is to understand how such factors shape writers' engagement with AI. \added{We present our results in relation to this earlier work, highlighting how our sample of experienced AI users extends existing insights. The values of \textit{intention}, \textit{authenticity}, and \textit{creativity} from Gero et al. in the context of social dynamics characterize when writers would choose to seek help from AI rather than a peer \cite{gero2023social}. Biermann et al. distinguishes the emotional values of \textit{fulfillment}, \textit{ownership}, and \textit{integrity}, in contrast with the value of  \textit{productivity} as competing motivations in writers' willingness to use AI. Building on this prior work, our study examines not only \textit{when} and \textit{why} writers choose to use AI, but also \textit{how} their values guide the way they use it. Prior challenges associated with AI such as maintaining writer voice, avoiding tropes, and understanding story structures \cite{ippolito2022creative} are somewhat mitigated in our study due to the advancement in model capabilities, the participants' experience with AI platforms, and the use of commercial platforms specifically designed for creative writing.}


\subsubsection{What motivates writers to use AI (and keep using it)?}
Though writers have varying motivations to incorporate AI into their writing practice, from creative expression to earning a living, all are ultimately determined to do what they love: writing. While some of our participants initially used AI out of curiosity, over time they uncovered the following benefits. 

\textbf{AI redefines relationships with creative blocks. }
The most common benefit was that AI transforms writer's block from an intimidating barrier into a manageable challenge. AI provided starting points as ideas or text, reducing the intimidation and cognitive load of figuring out what comes next, ``\textit{where with the AI, you don't have the mental game tripping you up anymore}” (P2). Writers don't always expect high-quality responses, but maintained confidence in their ability to shape any AI output into something useful. P13 adds: “\textit{I have a very low standard/expectation for AI’s generative responses, so I usually just take what it gives me, and I usually can run with that.}'' AI's ability to offer concrete ways to move forward acts as a ``safety net,'' helping writers maintain creative momentum and finish their writing. 

\textbf{AI can be a productivity boost and accessibility tool.}
AI can accelerate parts of the writing process, such as ideation and editing, where AI's constant availability shortens feedback loops. \added{Biermann et al. (2022) distinguish between hobbyist writers who resist AI to preserve emotional values and see AI as a tool, and professionals who prioritize productivity and see AI as a writing collaborator \cite{biermann2022fromtool}.} We find that the meaning of ``productivity'' varied among participants. For writers relying on writing as a source of income \added{(P2, P10, P12, P14)}, productivity is tied to maintaining competitive publishing speeds in an increasingly demanding market. For hobbyists, AI allows writers to complete projects otherwise constrained by time or energy \added{for which some welcomed AI help in translating their ideas to prose and still found ways to get fulfillment from expressing their ideas.} For writers managing health challenges, productivity shifted to enabling participation. P7, who struggles with carpal tunnel, uses AI to generate prose during first drafts: ``\textit{I do have flare ups... I was even considering hiring a co-author or ghostwriter... To have the AI do that really helped, really made me think that I can actually write again...}” P14, who writes for a living, mentioned AI as leveling the publishing playing field: “\textit{As someone who has had physical disabilities and the inability to work 40 hours a week for most of my adult life, suddenly I am competitive...}” 

\textbf{Writers can do more of what they love while developing their craft.}
Writers use AI strategically to reduce friction in challenging areas while maintaining focus on the parts of writing they find most fulfilling, creating a more satisfying creative practice. Each writer identified different parts of the writing process they enjoy or struggle with (such as plotting, writing dialogue, character descriptions). Writers, \added{especially hobbyists}, also noted that working with AI for challenging aspects became a learning opportunity, helping them develop their skills through example: ``\textit{this is a good learning experience for me too, because I'm not a perfect songwriter, I'm not a perfect poet. I don't think anybody is}'' (P13). \added{While prior work suggests that writers are more open to AI collaboration in areas where they lack confidence, our findings complicate that view \cite{biermann2022fromtool}. Writers in our study used AI strategically to reduce friction in parts of the process they found less enjoyable and not necessarily those they were less skilled in, though the two are often related. In fact, several experienced writers (especially in genre fiction and poetry) used AI in areas where they felt confident but felt to be tedious. Conversely, some hobbyist essayists avoided AI even in areas they struggled with, due to concerns about voice or authenticity.}

\textbf{AI enables new forms of creative experimentation.}
Writers deeply enjoy being creative, while still conceding “\textit{everything’s already been done before}” (P12). For many, creativity is instead more about the personal process of experimentation and shaping ideas in unique ways. Writers are looking for new ways of using AI beyond a basic writing helper and enjoy creatively experimenting in a low-stakes environment, \added{especially if they do not wish to use AI for prose generation directly.} 
For example, P3, \added{who enjoyed creating sentences and did not want to use any full sentences generated by AI} noted wanting \textit{``smaller, less fluent models producing random but cooler, wackier text... I would love to experiment with that.''} 
As another example, AI can help writers  venture beyond their usual writing styles and genres: ``\textit{within a lifetime, you can write more work, write more types of work, or you can be exposed to a lot of different genres really easily}'' (P17).

\subsubsection{When are writers cautious about using AI in their writing?}
Despite its benefits, writers are intentional about using AI. 
In order to preserve their values, writers set boundaries as part of their integration strategies by reducing AI use, limiting AI to specific tasks, or heavily revising AI output.

\textbf{Writers want to maintain their authentic voice, message, and sense of humanness.}
Writers approach authenticity through multiple lenses: their unique writers' voice and style, personal expression of message, and a sense of human presence in their work. Writers, especially experienced ones, have distinctive writing styles that both writers and long-time readers can quickly recognize, and will heavily rewrite AI text to maintain style or provide extensive text samples and prompts for AI to match.
In some cases, AI can help with intentionally exploring \textit{outside} of one's style: ``\textit{I got into AI to help with the parts of writing that I burn out on...to cover for the deficits I ran into... I don't do a ton of description in my writing. My brain just doesn't work that way, like my main character made it three books without a hair color... it's just not the type of stories that I want to tell, but this lets me indulge my childhood dream of being that type of person that can set a beautiful stage...}'' (P12)


However, when writers have a clear vision, AI becomes less helpful: they prefer to write directly without AI to achieve their intended message. The distinction becomes stronger with personal or reflective writing, where the process itself is cathartic or grounded in lived experiences  that ``\textit{AI can’t really dip into}.” Writers then prefer to leave AI for non-writing tasks like giving feedback, if used at all (P13, P17). 

\textbf{Writers want to retain a feeling of ownership and control.}
Writers locate ownership in different aspects of the writing process, leading to varying strategies for maintaining authorial control. For some, ownership centers on the initial creative vision. P2 explains, 
“\textit{That story premise was my idea. The words—AI wrote them, but I feel ownership of the idea.}” Importantly, writers adjust their AI use when it begins to detract from their enjoyment of the process. For instance, P2 offloaded too much to the AI \added{in trying to speed up the writing of a book} by using AI to create the main plot in a new genre and their sense of ownership degraded; P2 then decided to reduce AI use in their next project. While writers often want AI to fit their style, when it mimics their style well, writers may actually feel \emph{less} ownership. For P7:  ``\textit{it doesn't feel like 100\% me... 30\% to 35\% could easily be attributed to the AI, and I think part of that is because the AI is so good at mimicking the style and extending it}.''

Writers also find ownership through their editorial control and interpretation. P13 explains: ``\textit{I took that interpretation and made it my own}.''
Writers ultimately feel ownership through careful oversight of each decision: ``\textit{the AI couldn't get there without me, and the AI couldn't have done all those steps on its own}.''
They particularly emphasize needing control over the process given AI's lack of transparency: ``\textit{creatives always need to know what the prompt is}'' (P14). Many develop their own tracking systems, saving logs and maintaining version control to document their creative decisions. 

\textbf{Writers want to preserve their enjoyment of the craft.}
Respondents view writing as a site of engagement, not automation. They avoided using AI to generate text or ideas without first putting in their own effort or ensuring that significant editing and rewriting were involved. This commitment to the craft of writing spurs a reluctance to having AI do the work for them.
As P2 explains, ``\textit{my personal satisfaction is lower when the AI does most of the work...}''
P10 further states, ``\textit{If you weren’t a good writer writing stuff by hand, you’re not going to be a good writer with AI. It’s just going to extrapolate your foolishness}''. Understanding the writing process, genre expectations, and being able to distinguish and produce quality writing are important to writers. P6 notes that, for AI-assisted works, “\textit{you get to judge it on the quality. So if an AI wrote it… this guy was not involved in the process at all. He just clicked the button… it’s a story, but is it a decent story?}” Even when AI provides suggestions or generates text, writers believe they must still “\textit{put in the effort for it to come out with magnificent results}” (P6), applying rounds of revision and decisions to align with their vision.



\subsection{How Are Writers Incorporating AI into their Writing Process?}

As we've seen, writers carefully weigh AI's benefits against their core creative values, shaping \textit{how} they integrate AI into their workflows. What does that mean when it comes to the actual writing process? We find that writers maintain control through an iterative process.
Dynamic relationships with AI guide its use, and writers integrate AI at key decision points in new workflows.
At a high level, writers frame their relationships with AI, which can shift depending on the task. At a low level, we characterize the specific decision-making points in a writer's workflow when using AI. 


\subsubsection{How do writers determine their relationship with AI in practice?}

Writers perceive AI to take on a range of roles that directly shape \textit{how} they engage with it in practice and give attribution, influenced by their values and the writing task at hand. The writers in our study consistently drew on familiar human relationship metaphors such as ``assistant'' or ``collaborator'' to describe their interactions with AI, suggesting they conceptualize AI’s role through familiar relational dynamics rather than just functionality.

While AI might perform similar tasks across different relationships (both assistant and collaborator roles can derive plot ideas), how writers frame these relationships shapes their overall approach to working with AI. 
Below, we summarize \textit{writers' relationships with AI} and organize them based on their perceived power dynamics, from those where writers maintain explicit authority (boss/assistant, director/actor) to more collaborative exchanges (fellow writer, editor) to inspirational (muse), to mechanical framings (tool). \added{These roles were named by the writers, with their characterizations further informed by both the interviews and observed writing sessions.} We provide more example quotes for every relationship expressed by participants in Appendix \ref{sec:relationship}.

Several participants described AI in subservient roles.
As an \textbf{assistant}, AI is performing tasks 
given by the writer (decision maker and boss). For example, P12 sees AI as ``\textit{a very good assistant …kind of like a supportive yes man.}'' Similarly, as an \textbf{actor}, the writer is the director, guiding the AI to perform tasks or behave in certain ways that fit their needs and the vision they have in their heads. P6 says ``\textit{I think more my role in using this is kind of like a director, and he is the actor. So I give it what it needs, then I see what happens, and then, if I like it, I keep it going}.'' AI output is molded according to the writer's creative vision, much like an actor following a director's cues. When AI is the assistant or actor, writers maintain tight control and desire high quality output. 

In contrast, collaborative relationships akin to a fellow writer show greater receptiveness to AI's creative contributions, often emerging as writers develop better expectations for AI's capabilities. Writers describe AI as contributing ideas, writing text, and giving feedback in the capacity of another writer. These perceptions range from seeing AI as an active \textbf{collaborator}, ``\textit{bringing its own things to the table}'' (P16) and helping to shape the text alongside the writer, through ideas, prose, or feedback in an iterative manner, to more passive roles such as \textbf{junior writers}, with less experience and skill, needing significant input and editing from the writer to be useful. Also mentioned was a \textbf{ghostwriter}, whose role is to help the author get their ideas down in their voice, contributing substantially less to the creative direction. \textbf{Editor} roles represent a middle ground where writers maintain creative control but are open to collaborative feedback. 
Writers use AI to identify strengths and weaknesses, restructure text, and give suggestions. P13 uses AI to simulate ``\textit{the viewpoint of an editor for a writer before my essay gets published.}'' 

As a \textbf{muse}, writers can take inspiration from the AI, sparking creativity and helping them get unstuck. P10 notes that the inspiration sometimes comes unexpectedly where ``\textit{a lot of times it'll spark an idea for something else}'' (P10) other than the original goal. P16 describes AI as an ``\textit{interactive muse},'' highlighting a back-and-forth exchange that extends the traditional idea of a passive muse, lacking specific expectations and open to inspiration without dictating its form. 

Though a \textbf{tool} might imply mechanical assistance rather than creative contributions, those who viewed AI as a tool were not more likely to use AI for any specific tasks, but rather saw these contributions as another resource. Some writers saw AI as another tool in their toolkit, dependent on skilled hands, similar to Photoshop or Grammarly, for any part of the writing process: ``\textit{So it's this tool for me, and it's something that I use, but it doesn't replace what I want to do, and I don't think it can}'' (P1). 
Those who viewed AI as a tool were also less likely to want to give attribution to AI: ``\textit{I'd be hard pressed to give credit to the computer that isn't actually really involved in the decision making aspects...only responding the way it is based on what I give it}'' (P12).

\textbf{Writers have dynamic relationships with AI and are still negotiating AI's role in their writing.}
Writers' relationships with AI shift based on context, task, and emotional engagement.
Writers can perceive AI to have multiple different roles even within a single writing session.
For example, in one writing session, P2 saw AI as a tool, fellow writer, and editor. These relationships stem from both conscious choices and emergent experiences with multiple differing AI models and interfaces. 

Participants also sometimes struggled to articulate  AI's role, with a common sentiment being that AI is both a tool \textit{and} a collaborator, raising questions about how these roles can coexist. Some writers see AI as a tool, but their gut reactions tell them AI is contributing more than just mechanical output, complicating writers' understanding. P7 describes this tension: ``\textit{I don't want to say that the AI is alive... but I also don't want to say that it's this lifeless tool... so it's a strange, middle ground... it does feel collaborative... there is something over my shoulder helping me.}'' This ambiguity is partly attributable to the evolving and often unsettled public discourse surrounding generative AI. Writers often found themselves lacking the vocabulary to describe their interactions with AI fully within familiar relationship paradigms. As P11 describes, ``\textit{it's my friend, it's my muse, it's my bag of tricks, it's my cut up technique... It's someone to bounce with, someone to give me a little idea, to whisper in my ear...a tool with a face, or a tool with a soul.}'' These roles can also evolve over sustained use as writers gain experience with AI. P17 described their trajectory from first using AI as part of a college class curriculum: ``\textit{every week we were doing AI...I was like `This is so scary. I don't know what's gonna happen.' But then afterwards I realized that it's really a tool, and then I started to think about it more like a printing press.}"



\subsubsection{How are writers integrating AI into their writing process?}
\begin{figure*}
    \centering
    \includegraphics[width=0.7\textwidth]{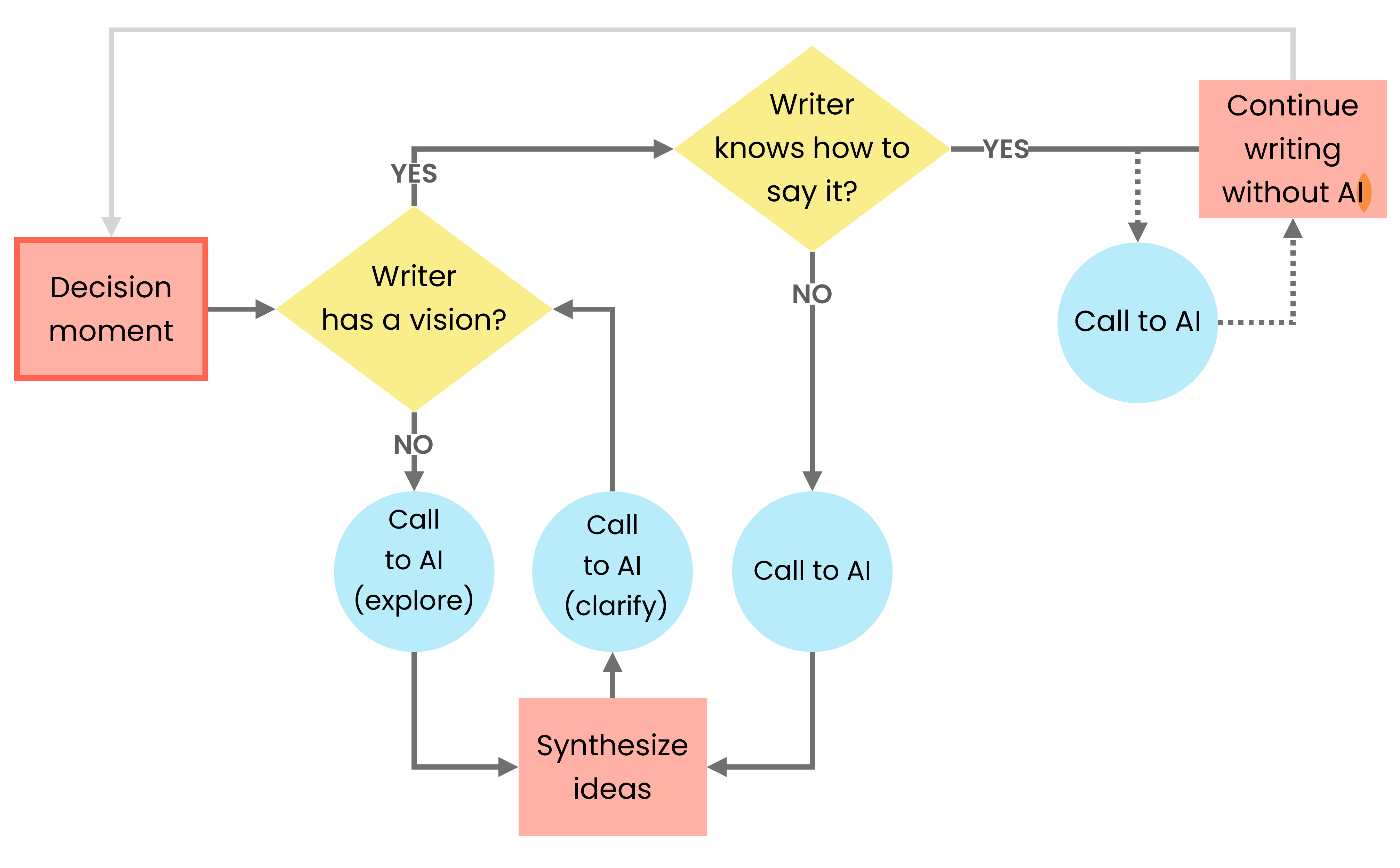}
    \caption{High-level interactions with AI during the writing process.}
    \vspace{-10pt}
    \caption*{\footnotesize Every time writers reach a decision moment (left-most box), based on their answers to key questions (yellow diamonds), they determine whether and how to call the AI (blue circles). Writers with a clear vision may use AI to help execute it or choose to continue writing directly. Writers without a strong vision of what they want may use AI to explore possibilities or clarify their ideas. The writing stages (red rectangles) show that AI calls are continuously synthesized towards a writers' vision. 
    }
    \label{flow}
    \hfill

    \Description{State diagram flowchart beginning with "Decision moment" leading to the question "Writer has a vision?" NO leads to "Call to AI (explore)" while YES leads to another decision "Writer knows how to say it?" NO leads to "Call to AI", while YES leads to "Continue writing without AI". Both calls to AI lead to the state "Synthesize ideas", which is followed by another "Call to AI (clarify), before returning to the decision "Writer has a vision?" }
    
\end{figure*}

\begin{figure*}
    \centering
    \includegraphics[width=.8\textwidth]{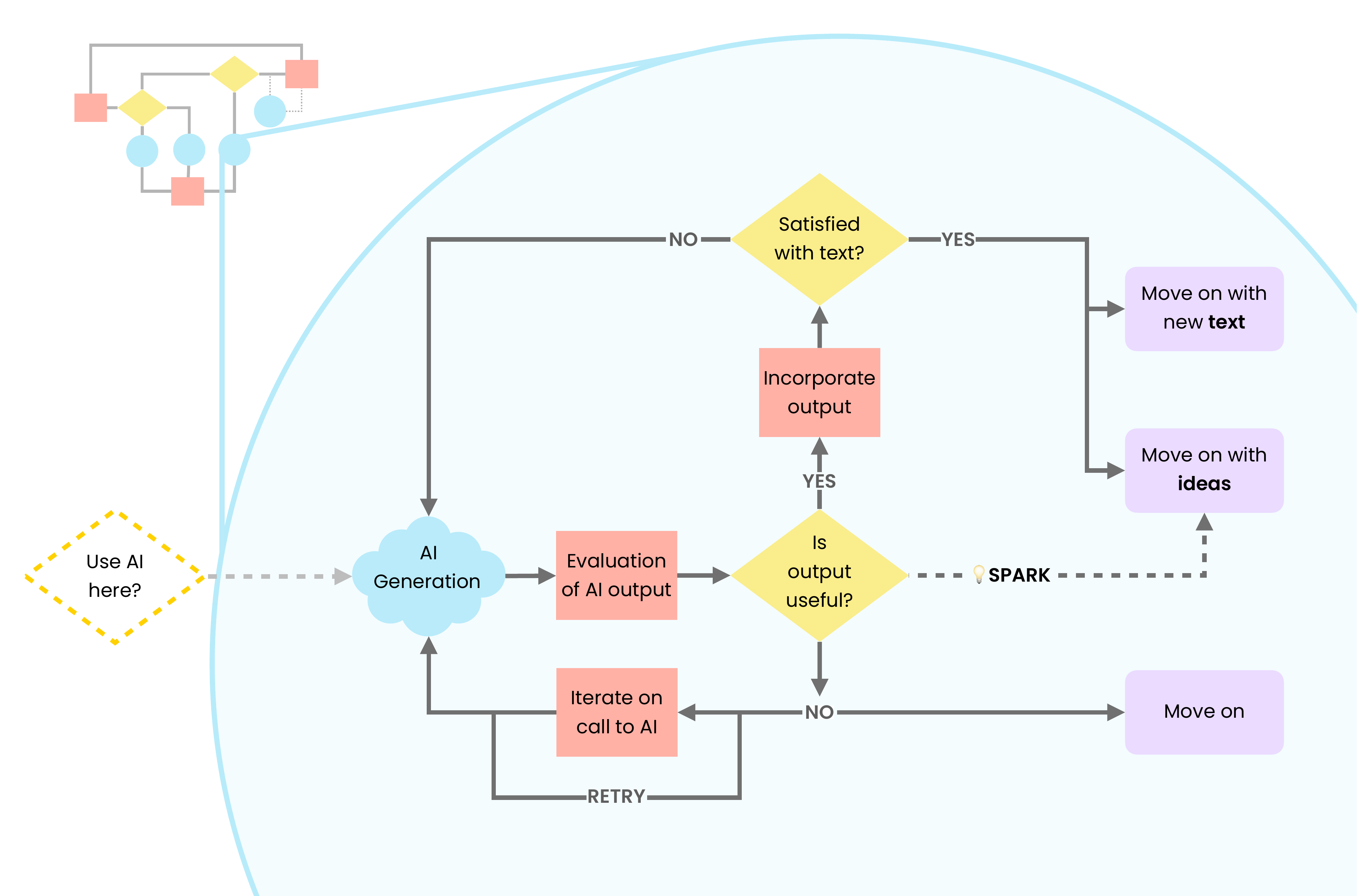}
    \caption{Detailed breakdown of writers' process each time they call the AI (corresponding to any ``Call to AI'' circle in Figure \ref{flow}).}
    \vspace{-10pt}
    \caption*{\footnotesize After AI generation (blue cloud), writers evaluate and make decisions about its usefulness. Writers can then iterate on the AI call or incorporate the output before deciding if the final text is satisfactory. AI can lead to surprising sparks of new ideas.}
    \hfill
    \Description{State diagram flowchart illustrating a zoom in on the "Call to AI" circle from Figure \ref{flow}. Outside this circle is the decision "Use AI here?" pointing inside the circle to "AI generation". After AI generation comes the stage "Evaluation of Output" leading to decision "Is output useful?". NO leads to either "moving on" and exiting the diagram or to the state "Iterate on call to AI", with an arrow pointing back to itself labeled "RETRY." These lead back to "AI generation." From "Is output useful?" YES leads to the state "Incorporate output", leading to the decision "Satisfied with text?" NO goes back to "AI generation" while YES lead to "moving on with text" or "moving on with ideas." Additionally there is a dotted connection from "Is output useful" labeled "SPARK" that leads to "moving on with ideas." }
    \label{zoom}
\end{figure*}

Writers integrate AI into their workflows in ways that reflect their creative values, goals, and relationships with AI. 
We summarize the \textit{integration strategies} writers deploy, such as how writers prompt, evaluate and incorporate text, in Figure~\ref{flow} and in more detail in Figure~\ref{zoom}. 
Overall, these strategies are not just technical processes but ways writers reconcile tensions, make choices, and experiment. 
A full list of all the tasks we uncovered for which writers use AI can be found in Appendix \ref{sec:AIuse}.

\textbf{Writers constantly make decisions around how to use AI in their workflow.}
Writers continuously make decisions about AI, from the initial setup (e.g., which platform to use) to individual writing moments (e.g., creating dialogue). \added{These decisions are shaped by prior experience and evolving familiarity with different AI, as writers develop a sense of which platforms and prompting strategies best serve their needs.} Beginning with the decisions around setting up their writing environment, writers noted that different AIs have distinct characteristics: some are better for prose generation, others more constrained and reliable for research, while some are ``crazy'' and better for creative ideation. For example, P1 uses ChatGPT for research and only uses Sudowrite for writing when stuck. \added{Seventeen out of 18 of the writers had used or started with ChatGPT as the most accessible entry point. As their needs for creativity, stylistic control, and consistency over longform texts grew, many began experimenting with more sophisticated prompts or branched into platforms with more specialized features for creative writing.}

Decision moments (Figure \ref{flow}) are key to understanding how writers integrate AI into the creative process at various levels of granularity, from individual words to entire plot lines.
During writing sessions, participants explained the moments where a writer might turn to AI. At any of these calls to AI, writers consider whether to use AI based on their vision and goals at each stage. When writers have a clear vision, they often choose to write directly to maintain their intended message. When their vision is less clear, writers may use AI to explore or clarify ideas while still maintaining editorial control over the final direction. \added{For example, P13 did not have a specific vision for their song but rather started with a general topic, using AI to surface relevant themes, characters, and settings. A phrase from this exploration struck their inspiration and ultimately became the song’s title.}
These decision points allow writers to adjust AI use based on their values and the current role AI is taking on. 
At any stage, writers may also shift from a focus on writing to revision, using AI to assess and refine their text with actions such as summarizing, improving structure and clarity, or aligning with genre expectations.
Writers might also turn to AI for enrichment, such as doing research or asking questions to deepen character motivations. 

\textbf{With every call to AI, writers also make many fine-grained decisions.}
At a more granular level, Figure \ref{zoom} maps the multiple sub-processes and decisions involved in an individual \textit{Call to AI} in Figure \ref{flow}. Writers decide whether to use AI for a specific task based on their goals and values, how to prompt effectively to get useful output, how to evaluate outputs against their standards, when to iterate versus move on, and how to incorporate useful elements. Practical considerations include whether writers are open to external ideas, there is enough existing text for AI to build on, or the task is one AI handles well. 
Writers' prompts ranged from complex, customized templates to conversational. 
Some writers describe prompts being as personal as their writing style, while others use the process of writing prompts to clarify their own ideas.
Generations also do not have to include a prompt, as some may be generated directly inline depending on the interface.

\textbf{Evaluation is where writers exert considerable control.} Writers evaluate AI output from multiple dimensions: alignment with their voice and vision, logical consistency, genre expectations, quality standards, and utility. Evaluation benefits from intuition built on experience in writing, as P14 notes in evaluating alignment with style: ``\textit{I can immediately evaluate a sentence as whether or not that's a sentence I have the ability to write... I'm a good writer to write with AI, because I've already done my internship, my apprenticeship, and my years in the trenches.}'' If outputs don't meet standards, writers iterate by adjusting prompts, modifying context, or providing additional information---until the writer receives something usable, and proceeds to incorporate the text, which often involves more edits on the part of the writer, and a closer inspection: ``\textit{I have read every single word and I am putting my stamp of approval that this can go out the door}'' (P14).

\textbf{Unexpected generations can still be useful.}
Even when AI produces something writers weren't aiming for, it can still trigger a ``spark'' \cite{gero2022sparks}---an unexpected moment of inspiration that pushes the writer forward. As AI generates more material, the chance of something catching the writer's attention increases. For P1: “\textit{just having something bad is useful, just to see how it flows.}” The process of generating and rejecting options also helps writers clarify their own visions, where even ``bad'' outputs serve as valuable \textit{anti-patterns}: showing writers what they don't want helps them understand what they do want. 
As P10 notes, ``\textit{even poor quality outputs are still useful for sparking ideas}.'' \added{Sparks often emerged unexpectedly: prompts that generated more text material to work with or surfaced a wider range of ideas were more likely to yield sparks (asking for themes (P13), putting characters in dialogue with each other (P1), or \textit{reverse-prompting}, where AI asked the writer questions back to reflect on their writing (P4)). For the genre of poetry especially, P16 explained ``\textit{all the mistakes, or the things that feel unnatural, or unexpected, they're immediately so much richer than something that feels familiar: anything that sounds slightly odd, like it sticks, kind of rings in the air. You want to work with it, and it suggests new ideas as well}.''}



\subsection{The Interplay of Values, Relationships, and Integration}

\begin{figure*}[h]
    \centering
    \includegraphics[width=0.75\textwidth]{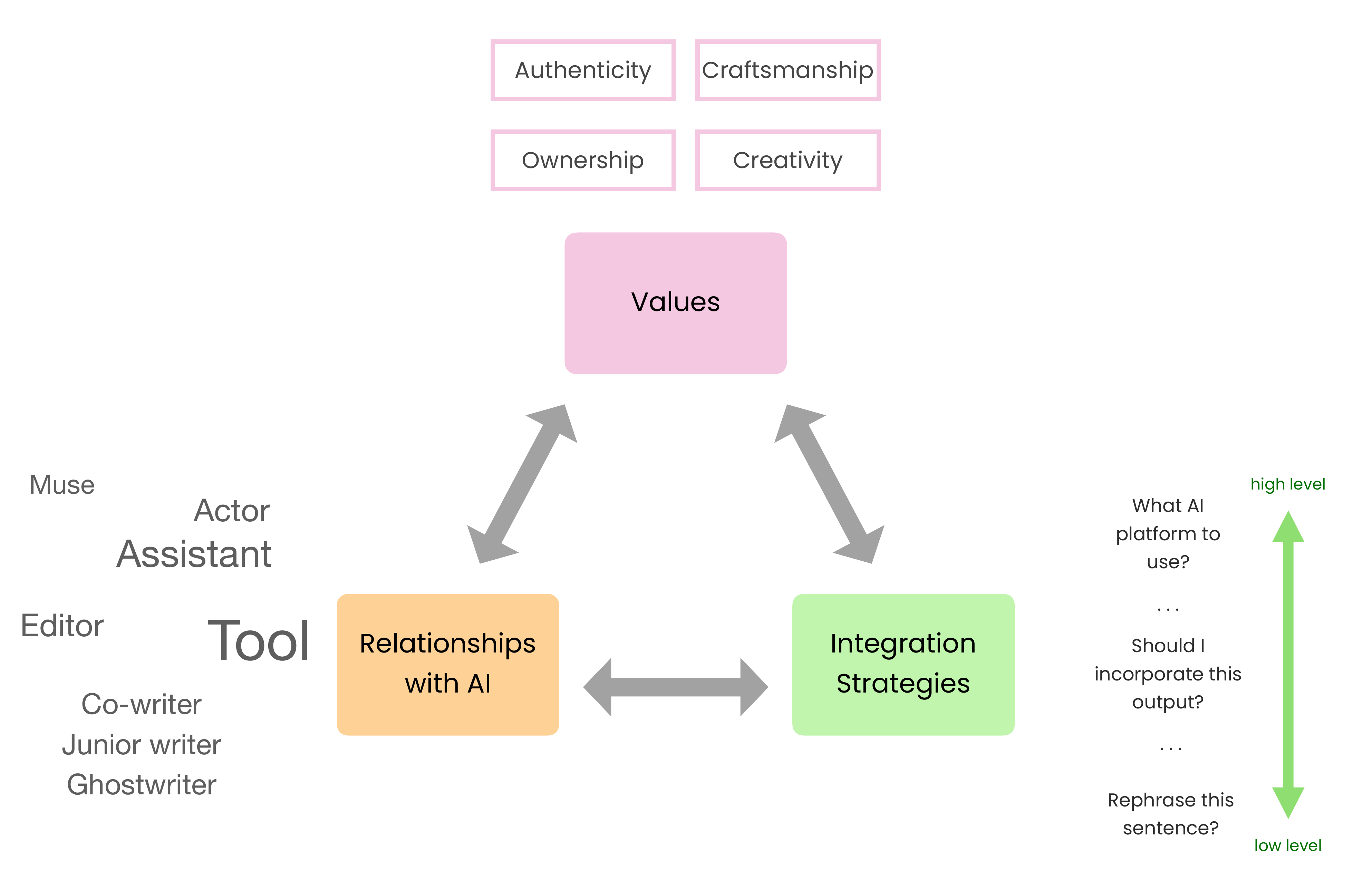}
    \caption{Three core elements that influence and reinforce each other as writers develop their AI writing practice: writers' values, their relationships with AI, and integration strategies.}
    \Description{Diagram with 3 double-sided arrows pointing between "Values", "Relationships with AI", and "Integration Strategies" arranged in a triangle. Next to values are the words Authenticity, Craftsmanship, Ownership, and Creativity. Next to Relationships with AI are the words Assistant, Actor, Co-writer, Junior writer, Ghostwriter, Editor, Muse, Tool in a floating word cloud. Next to Integration strategies is an arrow showing a spectrum from high level to low level use. 3 questions are written next to it. At the high level "What AI platform to use?". At the medium level "Should I incorporate this output?" At the low level "Rephrase this sentence?"}
    \hfill
    \label{summary}
\end{figure*}

The interactions between writers' values, their relationships with AI, and their integration strategies serve to explain observed patterns in writer behavior, where each influences and reinforces the others (Figure \ref{summary}). 

\added{\textit{Values and Integration Strategies.}} \deleted{Values set the foundation of what writers find most important in the writing process, the core beliefs and principles that guide behavior. }Writers choose when, why, and how to use AI in order to preserve and align with their values, from rewriting AI text to maintain authentic voice to creating experimental prompts for more creative exploration. Yet these strategies can also shift how writers experience their values over time and continued use. Writers can discover their initial boundaries (such as ``writing every word myself'' to preserve ownership) can evolve into more nuanced approaches as they use AI in new ways: \added{``\textit{I think that it's only reinforced my idea of originality and authentic self... you can absolutely be an authentic artist as long as it's your work and it's your interpretation}'' (P13).} On the other hand, when a writers' use of AI conflicts with their values, writers will adjust their strategies, reinforcing the importance of their boundaries, \added{as in the case of P2 reverting their  AI strategy to preserve feeling of ownership.}

\added{\textit{Values and Relationships with AI.}} Values, which are individual to each writer and their literary genres, also strongly influence how they first approach AI and conceive of its role. For example, those prioritizing ownership often begin by viewing AI as a tool, while writers from literary traditions more open to referencing and remixing more readily embrace AI as a muse or creative collaborator. However, these relationships can evolve to challenge writers' conceptions of their values, reconsidering what concepts like ownership and craftsmanship mean in practice, leading to further refinement of values. \added{P5, for example, began by seeing AI as a tool, but came to describe it as a “ghostly co-writer” with its own direction, prompting them to feel like an “imposter,” and ultimately reframe ownership as something shared rather than singular.}

\added{\textit{Relationships with AI and Integrations Strategies.}} How writers view their relationships with AI, whether as tool, collaborator, or muse, shapes their initial strategies for integration. Those who see AI as an assistant develop more task-specific workflows while those seeing AI as a muse might experiment more freely. \added{For instance, P8's view of AI as a human companion directly influenced their decision not to use any AI-generated text directly: ``\textit{I like to treat [AI] as a human being... and if I were to include any line from ChatGPT, I just feel like I can't credit it, so I won't use it.}''} The ways in which writers use AI in turn influence how writers understand AI's role. As writers discover new and effective approaches or as more sophisticated AI models arise, their conception of AI's role can expand. For example, many writers used ChatGPT and saw AI as an assistant due to its overly helpful tone and predictable responses, until they tried AI platforms for creative writing with fewer guardrails, which they then saw as more of a collaborator. Similarly, as writers run into challenges, their conceptions of AI as a limited assistant or tool can be reinforced.

These interactions and feedback loops help explain how different approaches to AI arise, including how extensively and where AI is used. At a high level, writers tend to fall along a spectrum: on one end are those who use AI primarily for supportive tasks around the writing---like research, feedback, and creative experimentation---while maintaining complete control over text generation (P1, P3--4, P8, P13, P18). On the other end are those who engage AI extensively for text generation, either working iteratively at the sentence level with careful refinement (P5--7, P11), or at a broader structural level generating and revising entire scenes or chapters (P2, P9--10, P12, P14).

These interactions also serve to explain some apparent contradictions we observed. For example, writers who express the same relationship with AI can develop widely varying workflows based on their underlying values and genre traditions. Similarly, writers who prioritize the same values can develop wholly different approaches due to their view of AI's role. For example, both P8 and P6 valued ownership, yet P8 refused to directly use AI-generated text while P6 did. For P8, ownership meant highly constrained usage of AI, while P6 felt ownership through their conception of themselves as a creative director of AI.

\section{Discussion}

Our findings reveal how creative writers are developing intentional approaches to AI that preserve their values while embracing the benefits. In this section, we examine the broader implications of these practices for the future of creative writing and the design and research of AI writing tools.


\subsection{Writers' AI Workflows}

The writers in our study demonstrate ways of working with AI that are far from automating the process, instead developing new workflows and expertise in areas such as prompting, evaluation, and creative experimentation. These new workflows suggest that rather than automating creative writing, AI may be creating new dimensions of writerly expertise. Writer workflows vary significantly across genres: poets often approach AI as a material generator to be reshaped, \added{being more comfortable in experimenting with prompts.} Fiction writers focus more on AI as productivity support, \added{for tasks like ideation and prose generation.} Personal essayists were most restrictive in not using AI to generate text, typically using AI only for research and feedback.
Rather than converging on a single model of AI use, we may see continued divergence as different writing communities develop practices aligned with their values and traditions.
This observation carries implications for new AI-infused writing tools: while scaffolded support may accelerate common tasks, it might also stymie more creative and genre-specific workflows.

The nature of these emergent behaviors can challenge traditional notions of creative work and automation. Unlike bounded tasks that can be automated, such as email summarization or report generation, creative writing serves as a \textit{boundless activity} where the work is never ``finished'' until the writer decides so---there are always possibilities for revision and new ideas. Our participants consistently had more ideas than they had time to write, using AI not to reduce effort but to enhance productivity and explore more. 



\subsection{Tool or Collaborator?}
Prior work has characterized AI writing support through categorical frameworks, such as tools for cognitive offloading or collaborative partners \cite{gero2019Metaphoria, Yang2022AIAA, redhead}. However, our findings reveal that writers develop more fluid relationships with AI that challenge these simple categorizations. Rather than trying to enforce rigid role boundaries, AI systems might better support writers by acknowledging and enabling fluid movement between different types of engagement. Writers often maintained multiple, simultaneous relationships with AI that resist clear classification as either tool or collaborator. 

\textit{Design and research direction: Expanding beyond familiar relational metaphors.} Writers drew on familiar human relationships, but found these insufficient to capture their actual practice, suggesting room for new metaphors to take root. AI systems' designs and capabilities are not neutral and influence perception of AI's role, such as ChatGPT's restrictions and tone being perceived as assistant-like. Future systems should explore opinionated designs geared towards new metaphors to expand what is creatively imaginable, such as a ``gardener'' that nurtures and curates their ideas, or a ``conductor'' orchestrating multiple AI functionalities at once.
Would would it look like for the writer to be a sculptor, starting with a mass of text and chiseling away to find the artistic form underneath? 


\subsection{Market Pressures, Productivity, and Creativity}

The tension between productivity and creativity in writing is not new \cite{Boice1993WritingBA, Kachelmeier2008MeasuringAM}. Professional writers, particularly in genre fiction where publishing speed impacts income, described feeling pressure to maintain competitive publishing speeds \added{with P10 stating \textit{``we get paid by the word, the more books I put out, the more money I make.}''} This suggests a potential future where AI adoption may become necessary for \added{competitive} market participation \added{in some genres}, rather than truly optional \added{especially as AI has been found to reshape job content as it amplifies productivity \cite{Eloundou, NBERw30074}.} Some writers turn to AI due to health challenges to level the playing field, raising questions about escalating productivity expectations. Our participants expressed ethical concerns such as environmental impact, models' training data source, and writer compensation, yet many balanced these pragmatically with productivity expectations.
Although all participants hoped for a future where AI would support human creativity, some participants predicted a future where AI-generated commercial writing and human writing would become increasingly separate: \added{``\textit{I think it's going to become like a craft art, if that makes sense. Like, for example, this shirt was woven on a machine loom...''} (P14).} Supporting sustainable creative careers will require new frameworks for valuing and compensating creative work that account for both human craft and technological assistance.

\textit{Design and research direction: Rethinking useful models for creative writers.} 
Our findings showed multiple ways of successfully using AI other than generating text, where writers often found value in ``bad'' outputs that sparked new ideas and used AI for experimentation. The observation that AI text does not need to be perfect or as complex as that produced by the largest language models invites investigation into smaller models, which may suffice for some writers' needs.
Focusing on smaller models makes easier the creation of specialized models focused on creative possibility and experimentation. 
It additionally makes easier the creation of consensually trained individual, or potentially community-driven models that can offer an alternative to larger platforms writers have little control over, assuaging the feeling of having to choose between productivity and personal concerns.

\subsection{Transparency of New Processes}


Unlike previous forms of computational support, where the technology makes the process apparent, disclosure of AI assistance reveals little about the actual creative process. Each writer in our study developed unique workflows and these individual approaches remain largely invisible unless explicitly documented. The uncertainty surrounding best practices highlights that many writers are still experimenting with how (or whether) to disclose their use of AI and the degree of its use. 
Writers were similarly unclear about how they would ideally give attribution to AI, such as attempting to approximate with percentage, give authorship credits, or acknowledge how and which AI was used.

Our participants developed various ad-hoc approaches to track their process (color-coding AI-generated text, copying prompt logs), but these personal systems do not easily translate to shareable knowledge. Most writers developed their workflows through trial and error, supplemented by writing communities and online resources. Current efforts around AI disclosure focus primarily on simple statements or surveillance \cite{GrammarlyAuthorship2024}, yet these approaches fail to capture the nuanced ways writers integrate AI into their practice. 

Additionally, broader societal stigma around AI-generated content make it even harder for writers to share their processes.
Despite how writers would ideally like to give attribution, whether they choose to disclose AI becomes a separate issue, due to stigma and perceptions of AI-assisted writing. 
Unfortunately, stigma was even internalized, as P5 said: ``\textit{I do feel like an imposter for using AI in the sense that maybe I'm cheating a little bit, or maybe I'm not working as hard as I could be}.''




\textit{Design and research direction: Supporting visibility of AI integration to share new processes.}
Rather than focusing on detecting or marking AI use, new tools should help writers document and optionally share their creative processes. Future interfaces could build in rich process documentation that enables writers to track decision points and creative choices in ways meaningful to them, serving both as personal reflection and potential knowledge sharing. These systems might support different levels of process sharing, from personal records to writing groups to public disclosure. The goal should not be surveillance, but supporting writers in understanding their own practice while enabling voluntary knowledge sharing within creative communities. This could help reduce stigma by moving discourse beyond binary AI or human debates toward more nuanced understanding of how writers maintain creative authenticity while using new tools. Over time, communities in different genres could develop new norms for AI attribution that extend from existing ones.
For instance, P15 shared they come from a poetic literary tradition with an ``\textit{unwritten code of ethics about referencing lines... It feels like poetry is very much a kind of a communal effort... So do I feel like [ownership] should be claimed?...it really depends on how the poem came about, and how heavily it leans on AI, and whether or not the knowing that AI was part of the process benefits the reading of the poem.}'' By making processes more transparent through voluntary sharing rather than policing, \added{future research could explore how such visibility helps communities develop best practices by revealing diverse workflows, fostering discussion, and modeling examples of responsible AI use.}

\section{Limitations}

We focused on writers already using AI in order to understand their workflows, which excludes the perspectives of non-users and AI opponents. Their underlying reasons and values for hesitation or opposition offer critical insights into the concerns around authenticity, creativity, and the broader impact of AI on the writing profession. While we recognize the significance of these views, they fall outside the scope of this particular study, which was designed to focus on the emergent patterns that arise out of AI use.

Our interview structure was shaped by the need to understand writers’ usual workflows, though we recognized that some participants might feel uncomfortable being observed while engaging in creative work. To mitigate this, participants could prerecord writing sessions or opt for a diary study. The first author also offered muted video during writing sessions. Overall, most participants reported comfort and enjoyment in verbalizing and sharing their process. A minority noted heightened self-awareness or rushed pacing (e.g., one participant spent less time between iterations). Most felt that the sessions reflected their typical practice or \added{explained how they might have done the writing slightly differently.}

\section{Conclusion}
This work examines how creative writers who have chosen to incorporate AI into their writing process view and use AI. We interviewed 18 writers across various genres, all of whom had experience using AI tools. We identify core elements that shape how writers engage with AI: their values, their perceptions of AI, and their strategies for use, along with the tensions that writers hold when using AI. Our findings reveal that AI’s role in creative writing is far from passive, requiring significant effort and decision-making by writers to align the tool’s output with their creative vision. These insights offer important implications for the design of future AI writing tools, emphasizing the need for customization, transparency, and support for creativity in ways that enhance the writer’s process rather than undermine it.


\bibliographystyle{ACM-Reference-Format}
\bibliography{main}


\appendix
\clearpage
\section{Interview Protocol}
\label{sec:interview_questions}
\added{Here we outline the questions that guided our interviews and the live-writing observation sessions. The order of questions asked was also adjusted depending on what topics arose naturally through conversation, or skipped if not relevant. In line with grounded theory methodology, the questions evolved over the course of the study in response to participant behavior, emerging themes, and the dynamics of individual conversations. We conducted two rounds of updates to the interview guide to elicit richer, more relevant material as new insights and patterns emerged. Many of these questions came up naturally through the course of the interview and then were officially added to the interview guide. These updated questions are marked with (R1) and (R2).}

\added{
\textbf{General Writing Background:} We asked a set of background and onboarding questions regarding participants’ writing practices and AI assistance in general.}
\added{
\begin{itemize}
    \item What kind of writing do you currently do, and for how long?
    \item What kind of formal or informal education have you had as a writer?
    \item What motivates you to write? Which aspect of your writing experience do you enjoy the most?
    \item How do you define success for your writing? (R1)
\end{itemize}}

\added{
\textbf{AI Background and Use:}}\added{
\begin{itemize}
    \item What AI-powered writing platforms have you used?
    \item What AI-powered writing do you currently use and how do you use them?
    \item How did you begin using these AI platforms? How did you learn to use them? (R1)
    \item What do you mainly use them for, and at what stage of your writing?
    \item When do you reach out for AI? At which stages? What are your goals for writing with AI and why?
    \item Which parts of using AI do you enjoy, and which parts do you dislike? (R1)
    \item How would you describe your core values as a writer? What is important to you in the writing process? Which of those feel supported or hindered by AI collaboration?
\end{itemize}}

\added{
\textbf{Live-writing Session:}
Participants were invited to live-write a creative piece using AI in their unique workflows while sharing their screens. Writers were encouraged to talk through the process but ultimately do what felt natural to them. During the live-write, we would occasionally ask questions about motivations and feelings, giving space for the writers to reflect and evaluate their interactions with AI. We began by asking for context on what they were planning to write and what their goals were.}
\added{
\begin{itemize}
    \item Why did you use AI here? / Why did you choose not to use it?
    \item How did you come up with that prompt?
    \item How are you evaluating the AI’s output?
    \item What were you hoping to get from the AI here?
    \item How do you feel about the results you got?
    \item What is the reason for doing X (action participant performed on screen)?
    \item Does it still feel like your idea/why do you feel comfortable using or not using the output?
    \item Does this feel like the final text? Will you come back and edit it?
    \item How strong of a plot or idea do you have when prompting the AI? Are you looking for ideas or looking to verbalize the vision you already have?
    \item After the session:
    \begin{itemize}
        \item How do you feel about the text that they wrote?
        \item How do you see attribution and ownership?
        \item Did you pick up on patterns or processes that you didn’t notice before? (R1)
    \end{itemize}
\end{itemize}}

\added{
\textbf{Walkthrough of Previous Writing:}
Participants were also invited to share a piece (or pieces) of writing they had previously finished with the assistance of AI writing tools. We asked questions about what they remembered about the steps of the process as well as reflections of their views on creativity, attribution and ownership, and important moments from integrating AI into their workflows.}
\added{
\begin{itemize}
    \item How long ago was this piece written? Do you remember how you used the AI?
    \begin{itemize}
        \item Which ideas were from the AI’s suggestions?
        \item Which words/phrases came from the AI’s suggestions?
    \end{itemize}
    \item Were there any memorable moments while writing this?
    \item What were some moments that you remember the AI being helpful or frustrating?
    \item How do you think about ownership and attribution in this piece?
\end{itemize}}

\added{
\textbf{Workflow Reflections:}
In this section, our follow-up questions focused on the role of AI tools in the different stages of writing. Our aim was not only to understand how these tools are being used but also to explore the reasons behind the non-use of certain AI functionalities, which might reveal underlying value conflicts. Specifically, we dived into their experiences of writing a creative text with AI assistance. While the live writing session provided insights into the detailed interactions during the actual writing of text, these conversations allowed us to explore the role of AI in the planning stage of writing, such as the planning and development of character personas and story plots. We also asked participants to share their perspectives on the value of AI written text, AI attribution in writing, and any ethical concerns.}
\added{
\begin{itemize}
    \item How would you describe your relationship with the AI or the role AI plays in your writing? Is there a point at which you would consider AI a co-author or collaborator? (R1)
    \item How has your relationship with AI changed over time? (R2)
    \item How do you think about disclosing your use of AI? (R1)
    \item What is your understanding of how these AI tools work?
    \begin{itemize}
        \item Has your view on AI changed over time as you've gotten more experience with it? (R2)
        \item Have your views on originality, creativity, and ownership evolved through working with AI?
    \end{itemize}
    \item Does your AI use vary across genres? (R1)
    \item What does an ideal future look like for AI and creative writing? (R1)
    \item Do you have any concerns about using AI?
\end{itemize}}

\section{Participants' AI use}
\label{sec:AIusesurvey}

These are the recorded answers from participants' pre-interview survey answering the question "Briefly, how do you use AI in your practice?" (Table \ref{tab:partcipant_ai_use_table}).

\begin{table}[!ht]
  \caption{Participant's pre-interview answers to the question "Briefly, how do you use AI in your practice?"}
    \centering
    \small
    \begin{tabular}{|c|p{0.9\linewidth}|} \hline
 \textbf{ID}&\textbf{AI use}\\ \hline 
 1&Topical research, organization, ideation, prose generation\\ \hline 
 2&Story ideation, create an outline, create character templates, create beats for chapters, write the prose, edit, ask question if this is in character, editing\\ \hline 
 3&Ideation, research, thesaurus\\ \hline 
         4& Grammar check, ideation, rephrasing, restructuring/flow\\ \hline 
         5& Ideation, rephrase, organization, stylistic auto-complete\\ \hline 
         6& Fiction writing and roleplaying\\ \hline 
         7& Prose generation, idea generation, research\\ \hline 
         8& I use it to give me promots (visual and text) or starting points as warm up excercises to kickstart my creative process\\ \hline 
         9& Ideation, generate poem with AI and editing\\ \hline 
         10& Ideation, prose, editing and proofreading\\ \hline 
         11& Bouncing my ideas, challenging me, probing/questioning, surprise factor, diving into the collective unconscious\\ \hline 
         12& Ideas, outlines, character profiles, writing the actual prose, suggesting beta feedback, etc.\\ \hline 
         13& I use AI to draw context from works I find inspiring so I can use this context in my own interpretation. \\ \hline 
 14&Ideation, refining ideas, writing long form passages, creating fine tunes on my own writing, editing, and marketing\\ \hline 
 15&Inspiration, rephrasing, summarization \\ \hline 
 16&As a muse! I ask AI chatbots to give me unusual images or facts, or to make contradictory statements. I also ask them to generate entire first drafts of poems, which I then re-draft extensively to make them more human. At the moment I'm especially interested in pushing back against AI bots' tendencies towards predictable or cliched content, and I'm experimenting with ways of prompting as well as ways of adapting AI-generated text that remove this predictability and make the text a bit more volatile and interesting\\ \hline 
 17&Brainstorm ideas, mitigate writing block\\ \hline 
 18&Varies considerably; one of the processes I am currently interested in is an iterative feedback process: I draft and ask GenAI for feedback, then redraft, and so forth. In other projects, the GenAI has been prompted to produce material, which may be revised, used as content, or used as a sort of prompt or launchpad.\\\hline
    \end{tabular}
  
    \label{tab:partcipant_ai_use_table}
\end{table}

\section{Summary of AI Platforms}
\label{sec:AIplatforms}

\added{Here we provide an overview of the most common AI writing platforms that participants used in this study. For each platform, we highlight the distinctive interface elements and main features (not exhaustive) that participants used in their creative process.}

\subsection{General Purpose Conversational AI}

\begin{figure}[h]
    \centering
    \includegraphics[width=0.45\textwidth]{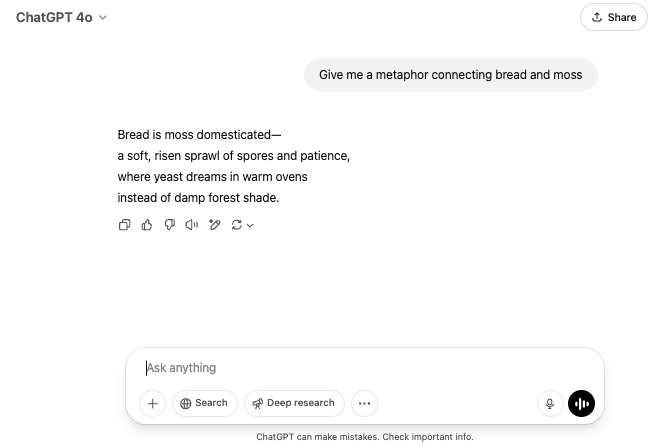}
    \caption{ChatGPT interface showing a chat interaction.}
    \Description{Screenshot of ChatGPT interface. The user asks 'What is a metaphor connecting bread and moss?' ChatGPT responds below that with an initial metaphor 'Bread is moss domesticated.' The interface shows a text input box at bottom.}
    \hfill
    \label{chatgpt}
\end{figure}

\added{ChatGPT (Figure \ref{chatgpt}), Claude, Copilot, Perplexity and Gemini are all general purpose conversational AI platforms (not made for creative writing specifically). All use a chat-based interface where users type prompts or upload documents and receive text responses in a back and forth manner.}

\subsection{Specialized Creative Writing Platforms}

\added{Unlike general-purpose conversational AI, several platforms are designed specifically for creative writers with features tailored to fiction writing, poetry, and other creative genres. These specialized platforms (NovelCrafter, NovelAI, Sudowrite, LAIKA, and Rexy) typically offer inline text generation that integrates directly with the writing process , along with custom AI models and unique features for different aspects of the writing process. These platforms combine AI assistance with organizational tools that writers can use independently of the AI features, such as support developing story plans and keeping track of plot points and characters. Below we summarize the common AI features present in many of these platforms.}

\begin{figure}[h]
    \centering
    \includegraphics[width=0.45\textwidth]{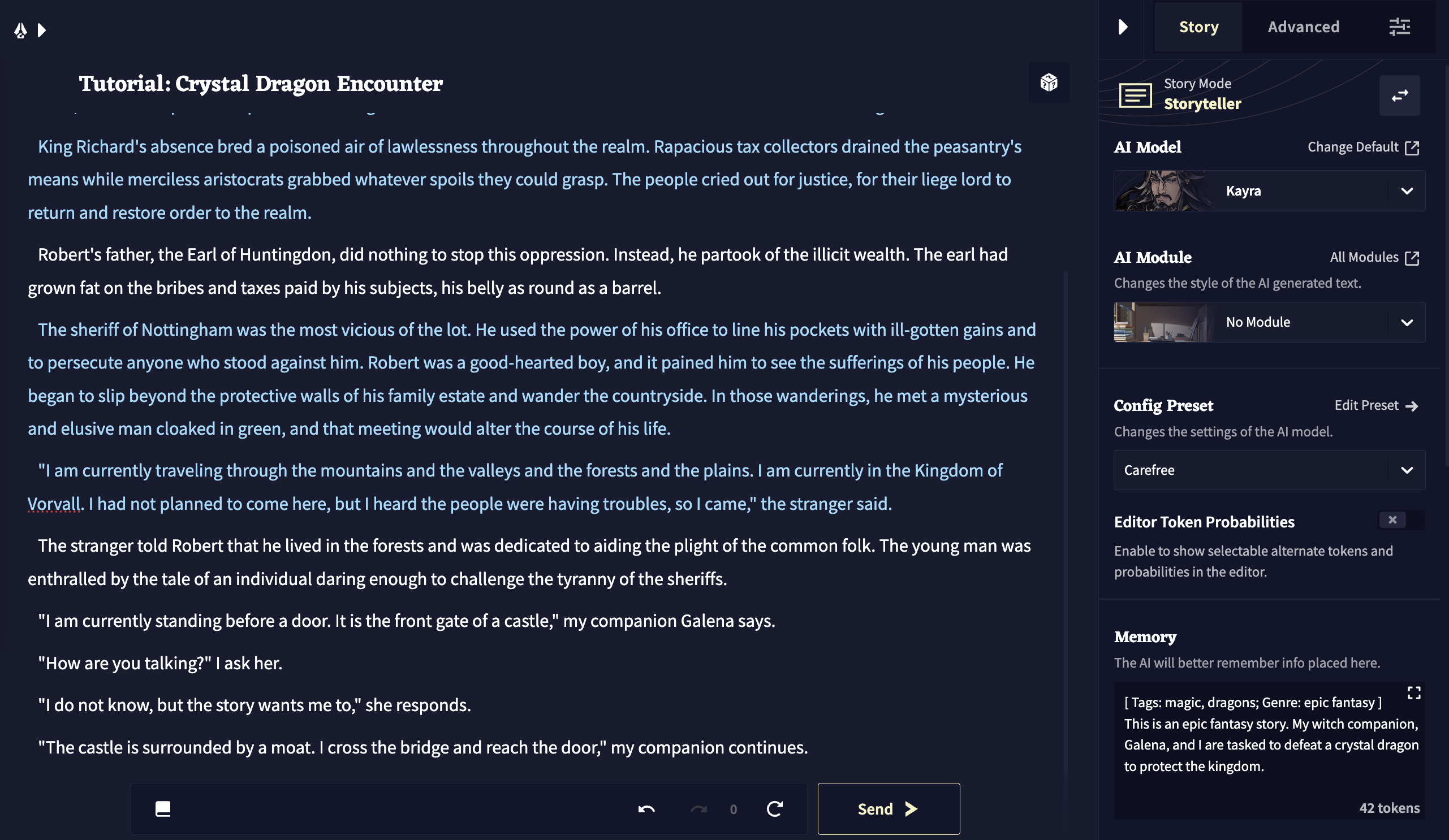}
    \caption{NovelAI Interface, showing AI writing in white and human or human edited writing in blue. The panel on the right shows a few of the model settings.}
    \Description{A screenshot of NovelAI's writing interface showing the tutorial chapter draft with prose about a king and his kingdom. A sidebar on the right shows model settings, including the AI model name: Kayra, AI Module (changes the style of the AI generated text): no module selected, Config Preset (changes the settings of the module: carefree, and Memory (the AI will better remember info placed here) in which there is the following text "This is an epic fantasy story. My witch companion, Galena, and I are tasked to defeat a crystal dragon to protect the kingdom."}
    \hfill
    \label{novelai}
\end{figure}

\begin{figure}[h]
    \centering
    \includegraphics[width=0.45\textwidth]{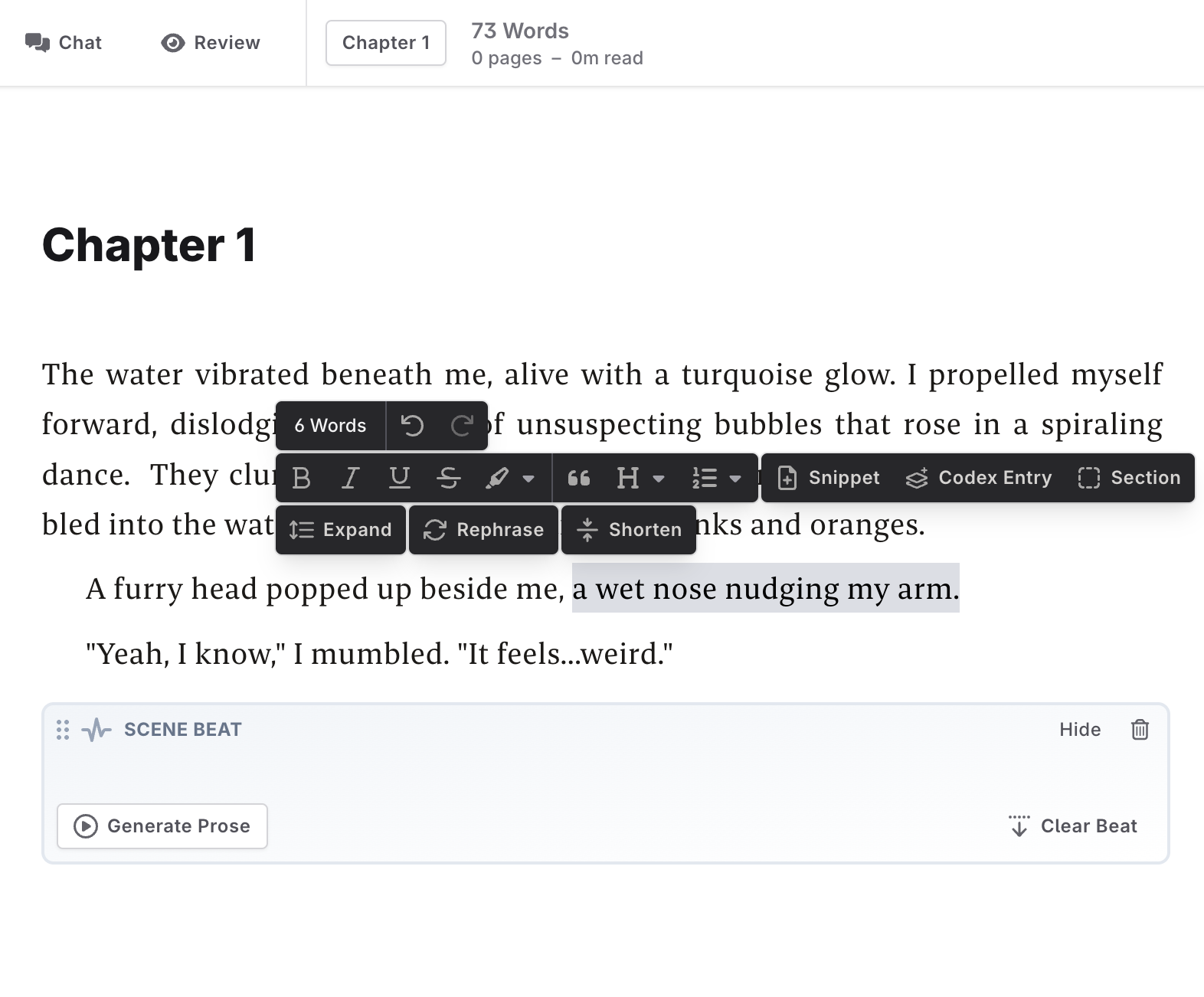}
    \caption{A portion of Novelcrafter's drafting interface. The SCENE BEAT feature allows writers to outline the scene for AI to generate. The toolbar shows options for the selected text, such as expanding, rephrasing, or shortening.}
    \Description{A screenshot of Novelcrafter’s drafting interface displays a draft of Chapter 1 with prose describing an underwater scene: 'The water vibrated beneath me, alive with a turquoise glow... A furry head popped up beside me, a wet nose nudging my arm.' Below that is an empty text box labeled SCENE BEAT with a button labeled "Generate Prose." The text "a wet nose nudging my arm" is selected and highlighted. Above that selection is a toolbar with text editing options and buttons labeled Expand, Rephrase, Shorten.}
    \hfill
    \label{novelcrafter}
\end{figure}

\begin{itemize}
    \item \added{\textbf{Inline text generation}: Writers can place their cursor in the middle of a paragraph, and the AI will generate a continuation directly in place. These can range anywhere from a sentence in length to a few paragraphs. 
    \item \textbf{Scene or chapter scaffolding}: Writers can provide a short summary, a list of plot points, or a few “beats” (key story moments), and the AI will generate a draft of a full scene or chapter based on that outline (Figure \ref{novelcrafter}.}
    \item \added{\textbf{Story and worldbuilding tools}: Writers can create detailed character profiles, track plot timelines, and maintain notes on fictional settings. Examples are Novelcrafter’s Codex: a manual, wiki-style database for structured character profiles, settings, and timelines, and NovelAI’s Lorebook: a repository for information about the story's characters, settings, and other lore that is added to the AI's context.}
    \item \added{\textbf{Model support}: Some platforms allow users to choose between different AI models (like GPT-4, Claude, custom or fine-tuned models shared in the community) based on their writing preferences, as well as customize the model parameters. LAIKA featured "brains": AI writing partners trained on either your own writing, or public domain texts.}
    \item \added{\textbf{Prompt customization}: Rather than starting from scratch each time, writers can save and reuse custom prompts and get support in creating prompts.}
    \item \added{\textbf{Editing and rewriting support}: AI can help rewrite a passage to improve tone, make it more concise, or shift its style. Some tools even offer sentence-by-sentence suggestions or allow writers to highlight specific parts for revision.}
    \item \added{\textbf{Color-coded annotations}: Some tools allow writers to mark which parts of the text were generated by AI, and which parts were written or edited by the author (Figure\ref{novelai}).}

\end{itemize}

\section{Writers' Values}
\label{sec:writers_values}
We identify four core values that creative writers want to preserve while using AI: authenticity, ownership \& control, creativity, and craftsmanship. The descriptions provided for each value aim to distill what writers mean when they refer to these concepts---while noting that participants often interpreted these values in different ways (Table \ref{tab:values_table}).

\begin{table}
\centering
\caption{Writers' values codes and example quotes.}
\footnotesize
\begin{tabular}{|p{0.18\linewidth}|p{0.25\linewidth}|p{0.43\linewidth}|} \hline
\textbf{Value} & \textbf{Description} & \textbf{Quotes} \\ \hline 
\textbf{Authenticity} & Importance of writing feeling that the writing comes from the author. & \textit{(After AI generating a few lines)} "But beyond that, that's when I start getting a little bit uncomfortable with how much am I losing myself in this writing process to something that's not me authentically.'' ---P13

\vspace{0.2cm}

``Keeping it still sounding like me is, like, really important to me.'' ---P4\\ \hline 
\textbf{Ownership \& Control}& Importance of feeling like the author has control and has ownership of the writing. & ``Having ownership of what I've written does mean a lot to me, actually, especially if it's something very personal.'' ---P13

\vspace{0.2cm}

``I would never copy-paste straight from\ldots\ the output that I get from ChatGPT\ldots\ I would still\ldots\ feel a sense of ownership over it.'' ---P4\\ \hline 
\textbf{Creativity} & Importance of writer feeling that they express themselves creatively in a way that makes their work distinct. & ``So I want to make sure, if I write a character, that the character's approach\ldots\ is unique, and it kind of challenges the reader to be like, `Huh? I've never thought of whatever topic in that way before.'\,'' ---P6\\ \hline 
\textbf{Craftsmanship}  & Importance of the effort and process that goes into producing quality writing. & ``So I can either invest the next year and learn how to write really exact beats so that it never does that, maybe. But I'd just rather write. I just want\ldots\ to get done, and I don't want to write AI-generated books that don't do anything.'' ---P1

\vspace{0.2cm}

``This is not a you push a button and a book falls out; you still gotta work at it. You still have to know your craft. If you weren't a good writer writing stuff by hand, you're not going to be a good writer with AI. It's just going to extrapolate your foolishness.'' ---P10\\ \hline
\end{tabular}
\label{tab:values_table}
\end{table}

\section{Participant Writing Session Setup}
\label{sec:writingsession}

\added{During the observed live-writing sessions, participants engaged in their writing practice as usual. Some worked on continuing in-progress pieces while others started on new projects. We summarize the writing task they worked on, how they used AI during that time, the specific AI used, and notes on how they set up their writing tools (Table \ref{tab:observation_table}).}

\begin{table*}
\centering
\caption{Live-writing session observed activity: summary of what each writer worked on during their writing sessions along with how AI was used.}

\footnotesize
\begin{tabular}{|c|p{0.13\linewidth}|p{0.31\linewidth}|p{0.08\linewidth}|p{0.20\linewidth}|p{0.14\linewidth}|}

\hline 

\textbf{ID} & \textbf{Writing task} & \textbf{Writing session activity} & \textbf{AI used} & \textbf{Notes on setup} & \textbf{Writer's View of AI}\\
\hline
1 & Chapter of thriller novel & Using AI to do research and scene development. & ChatGPT & ChatGPT and writing document in separate tabs & Tool \\ \hline 
2 & Chapter of romance novel & Using AI to reword plot points, generate prose, and then manually editing and writing. & Novelcrafter & Workfing fully in Novelcrafter, keeping notes of custom prompts in Notion & Tool, Director/Actor \\ \hline 
3 & Short story & Using experimental custom GPT to take fragments of ideas to combine into new story. Editing existing section, using AI to flesh out motivations of characters. & ChatGPT, custom GPTs & ChatGPT and writing document in separate tabs & Tool, Assistant \\ \hline 
4 & Personal essay & Using AI to edit, reword, and get feedback. & ChatGPT & ChatGPT and writing document in separate tabs & Tool, Editor \\ \hline 
5 & Sci-fi short story  & Using AI to generate inline continuations, choose from a branching interface, then manually continuing to write. & code-davinci-002 & Plugin that generates inline to the writing document (a branching text interface) & Collaborator, Tool \\ \hline 
6 & Fanfiction  & Generate continuations inline and editing to correct logic, manually writing to direct dialogue and pacing. & NovelAI & Working fully in NovelAI & Tool, Director/Actor, Ghostwriter \\ \hline 
7 & Chapter of fantasy novel & Using AI to generate inline continuations. Editing and writing to guide AI toward specific story vision. & NovelAI & Working fully in NovelAI & Director/Actor, Collaborator \\ \hline 
8 & Personal essay (starting new) & Using AI to do research, and explore themes and styles. & ChatGPT & ChatGPT and writing document in separate tabs & Tool, Collaborator \\ \hline 
9 & Poem (starting new) & Using AI to generate a love poem, then manually editing. & ChatGPT & ChatGPT and writing document in separate tabs & Tool \\ \hline 
10 & Romance novel & Using AI to reword plot points, generate prose, and then manually editing and writing. & Novelcrafter & Writing in Novelcrafter, Perplexity for ideation & Tool, Junior Writer, Assistant \\ \hline 
11 & Short story (starting new) & Using AI to generate suggestions inline, chatting with personals for plot twists, new interpretations, character crafting. & LAIKA & Writing directly in Laika & Tool, Muse \\ \hline 
12 & Chapter in new novel of a romance series & Editing Novelcrafter generated prose in Claude to match writer vision. & Claude,  Novelcrafter & Back and forth between Novelcrafter and Claude, keep track of ideas in Notion & Co-writer, Assistant, Tool \\ \hline 
13 & Song lyrics (starting new) & Using AI to generate themes and ideas, and restrcuture verses. & ChatGPT & ChatGPT and writing document in separate tabs & Tool, Collaborator \\ \hline 
14 & Romance novel (starting new) & Using AI to generate outline, edit outline, and generate prose. & Rexy & Using Rexy to write in Notion & Tool, Director/Actor \\ \hline 
15 & Poem (starting new) & Prompting ChatGPT to produce material on a theme, then collecting phrases to arrange into a poem. & ChatGPT & ChatGPT and writing document in separate tabs, collects by color & Tool \\ \hline 
16 & Poem (starting new) & Using AI to generate poem on a theme, then editing with further prompts (interpretation of the poem, new phrasings, new metaphors). & ChatGPT & ChatGPT and writing document in separate tabs & Muse, Tool, Collaborator \\ \hline 
17 & Song lyrics (starting new) & Using AI to generate lyrics from custom outline and prompt structure, then manually editing and rewriting. & ChatGPT & ChatGPT and writing document in separate tabs & Tool \\ \hline 
18 & Poem & Starting from found material (exisiting text) using AI to brainstorm ideas, then get feedback on drafts and advice on the genre. & Claude, Copilot & Claude/Copilot and writing document in separate tabs & Tool, Editor \\ \hline

\end{tabular}
    
    \label{tab:observation_table}
\end{table*}

\clearpage
\clearpage

\section{Writers' Relationships with AI}
\label{sec:relationship}


\begin{table}[!ht]

\caption{How writers see AI's role in the writing process.}

\centering
\renewcommand{\arraystretch}{0.9} 
\small 
\begin{tabular}{|p{4cm}|p{12cm}|}

\hline
\textbf{AI Role (Writer Role)} & \textbf{Quotes} \\
\hline

\textbf{Assistant (Boss)} & ``I come up with the characters, I come up with the storylines, and have the AI help me flesh that stuff out as an assistant.'' ---P10 \\ \hline

\textbf{Actor (Director)} & ``I'm like a strict director, you know? It's like, I have a vision, and I want to get to that vision, you know?'' ---P7 \newline
``So it's a lot like being the director of a movie. So the actors are the ones who are acting and all that. But I can be like, cut, do another take.'' ---P14 \\ \hline

\end{tabular}

\begin{tabular}{|p{1.8cm}|p{1.833cm}|p{12cm}|}

\multirow{9}{*}{\textbf{Fellow Writer}} & \textbf{Co-writer/\newline Collaborator} & ``It's a fellow author I can bounce ideas off anytime.'' ---P2 \newline
``I think it is a collaborator as well, but I have ultimately had the final say on what goes on the page and what doesn't.'' ---P3 \newline
``I probably go the collaborator route for most of it, because that's usually what I'm doing, is, okay, like, here's the problem we've got. How, like, can you help me solve this?'' ---P12 \newline
``To me, that's a collaboration, versus when I go to, like, Perplexity, and I'm like, 'Okay, I need this list of things that you can Google,' to me that is not a collaboration, because it's not, it's not like an iterative thing.'' ---P12 \\ \cline{2-3}

& \textbf{Junior Writer\newline (Senior Writer)} & ``Group of junior writers who are super eager and super crappy sometimes to just do whatever you want to do.'' ---P14 \newline
``You're the head writer. You may have writers that write under you, but ultimately you're responsible for the story and the final product. And that's the kind of relationship you have to take with AI.'' ---P10 \newline
``I would say it's a very eager junior co-writer, and it is a junior co-writer because it really can't do anything without my direction. And I would definitely say it's subordinate to me.'' ---P14 \\ \cline{2-3}

& \textbf{Ghostwriter} & ``So it's basically like having an electronic ghostwriter do your first draft, and then I'll go back and edit.'' ---P10 \newline
``It feels more like a ghostly co-writer, almost like a voice, rather than anything really substantial.'' ---P5 \newline
``So I'm basically creating a ghostwriter, but the ghostwriter is me, but I'm not writing.'' ---P6 \\ 

\end{tabular}

\begin{tabular}{|p{4cm}|p{12cm}|}
\hline
\textbf{Editor} & ``AI is like an editor that I can contact at any time.'' ---P2 \\ \hline

\textbf{Muse} & ``It's sort of like an interactive muse or something like that.'' ---P16 \newline
``AI is my muse, my bag of tricks, someone to bounce ideas with, to whisper in my ear, and keep me going.'' ---P11 \\ \hline

\textbf{Tool} & ``AIs are tools. It's literally a Swiss Army knife. It is a hammer, a wrench, screwdriver, all in one. It all depends on how you use it, you know.'' ---P6 \newline
``Pieces that I have done without AI and with AI. So I think AI is just another tool in my writing process.'' ---P5 \newline
``Like, it's just like a tool that I leverage in the same way that I use Grammarly.'' ---P4 \newline
``I see it as a tool, just like Adobe Photoshop. Like it's a tool.'' ---P2 \\ \hline

\end{tabular}
\label{tab:roles}
\end{table}

\clearpage


\section{Writers' Specific Uses of AI}
\label{sec:AIuse}

\begin{table}[H] 

\caption{Common AI use during observed writing sessions.}
\centering
\renewcommand{\arraystretch}{0.9} 
\footnotesize 
\begin{tabular}{|p{2cm}|p{3cm}|p{10.7cm}|} 
\hline
\textbf{Writers' Use of AI}& \textbf{Category} & \textbf{Actions} \\
\hline

\multirow{14}{2cm}{Prompting for Ideas and Inspiration} & Generating ideas & Pure continuation generations to see where AI goes next \\ \cline{3-3}
& & Prompting AI to enrich characters and create plot twists \\ \cline{3-3}
& & Searching for relevant themes to integrate in writing \\ \cline{3-3}
& & Generating multiple ideas at once and comparing different versions \\
\cline{2-3}
& Gaining sense of direction & Generating ideas for writing with an open mind and no vision \\ \cline{3-3}
& & Using AI as a checkpoint on whether to continue down a path of writing \\ \cline{3-3}
& & Requesting an outline/structure based on ideas \\
\cline{2-3}
& Overcoming writer's block & Generating starting points as a springboard to get unstuck \\ \cline{3-3}
& & Using AI to generate placeholder content\\ \cline{3-3}
& & Prompting AI with fragments of ideas to generate new output \\ \cline{3-3}
& & Using AI output as a way to reflect and get unstuck, even if it is not what the writer is looking for \\
\cline{2-3}
& Discovering sparks of inspiration & Using fragments of AI generations to write something new and usable instead of using exact output \\ \cline{3-3}
& & Prompting AI to emulate a specific writing style for inspiration \\
\cline{2-3}
& Putting ideas on the page & Prompting AI to help with the flow of ideas rather than prose \\ \cline{3-3}
& & Providing AI with the writer's intentions to generate ideas according to their vision \\
\hline

\multirow{3}{2cm}{Researching Content} & \mbox{} & Prompting AI to research a specific topic or find out a fact \\ \cline{3-3}
& \mbox{} & Asking AI follow-up questions to gain more in-depth knowledge and context about the topic they are writing about \\ \cline{3-3}
& \mbox{} & Using AI as a search engine to research, similar to using Google \\
\hline

\multirow{6}{2cm}{Planning the Writing Process} & Working towards a vision & Asking AI how to connect plot points and bridge different ideas \\ \cline{3-3}
& & Generating content to overcome the challenge of turning ideas and vision into a tangible outcome \\ \cline{3-3}
& & Generating new chapters based on writing style and outline, letting AI fill in the gaps \\ \cline{3-3}
& & Prompting AI to iterate and refine plot ideas for the story \\
\cline{2-3}
& Expanding words & Expanding specific aspects of AI-generated output by prompting \\ \cline{3-3}
& & Generating text from beats, which are outlines of scenes \\
\hline

\multirow{13}{2cm}{Revising and Generating Feedback}& Asking specific questions and seeking advice & Asking AI for feedback to broaden audience \\ \cline{3-3}
& & Requesting AI feedback on how to improve writing \\ \cline{3-3}
& & Asking AI questions about the flow of the text and if it makes sense \\ \cline{3-3}
& & Asking AI questions that inform the motivations and behaviors of characters \\ \cline{3-3}
& & Asking AI for perspectives to ensure the writer's ideas are conveyed clearly \\ \cline{3-3}
& & Using AI to benchmark whether writing is perceived the way the writer intends \\
\cline{2-3}
& Finding the right words & Prompting AI to flesh out social interactions in the story \\ \cline{3-3}
& & Using AI-generated snippets and synthesizing them to make writing more cohesive \\ \cline{3-3}
& & Prompting AI to refine specific word choices and reword sentences \\ \cline{3-3}
& & Generating words that rhyme and asking for synonyms \\
\cline{2-3}
& Following up after generating AI content & Refining AI generations by rewriting, adding elements of humanism, adding more details, and editing to match a specific style \\ \cline{3-3}
& & Iterating closely on each sentence generated by AI and iteratively using AI for writing the entire story \\ \cline{3-3}
& & Editing AI output to maintain logical consistency in writing and adjusting AI prompts to reestablish direction and vision \\ \cline{3-3}
& & Pure continuation generations to continue the flow of AI writing and see where it goes next \\ \cline{3-3}
& & Writing to provide more context before prompting AI and regenerating content \\ \cline{3-3}
& & Researching specific topics using AI to gain more domain knowledge and delve into specific topics \\ \cline{3-3}
& & Regenerating content when previously generated content is not what the writer is looking for \\
\hline

\end{tabular}
\end{table}




\end{document}